\documentclass[aip,pof,reprint,amsmath,amssymb,floatfix]{revtex4-2}

\usepackage{graphicx}
\usepackage{booktabs}
\usepackage{url}

\newcommand{\Fr}{\mathit{Fr}}
\newcommand{\We}{\mathit{We}}
\newcommand{\St}{\mathit{St}}
\newcommand{\Cpn}{C_{pn}}
\newcommand{\Cpmax}{C_{p\max}}
\newcommand{\dd}{\mathrm{d}}

\begin{document}

\title{Annular liquid jets: exact constraint absorption and finite-time
loss of global admissibility}

\author{Francisco R. Villatoro}
\email{frvillatoro@uma.es}
\affiliation{Escuela de Ingenier\'ias Industriales, Universidad de
M\'alaga, 29071 M\'alaga, Spain}

\date{\today}

\begin{abstract}
An annular liquid jet may enclose a gas volume and close on the symmetry
axis at a finite convergence length. In the membrane limit, closure supplies
an algebraic relation between the state variables at the moving free end.
Earlier domain-adaptive schemes differentiated this relation to obtain an
evolution equation for the domain length. This differentiation is exact at
the continuous level as long as the closure relation is satisfied. In a
discrete calculation, however, integrating only the differentiated relation
can allow the numerical solution to drift away from the original closure
condition unless projection or stabilisation is used. For the membrane
problem ($\beta=0$), we avoid this source of drift by writing
$R=(1-\eta)S$, which satisfies the endpoint closure identically, and then
apply Chebyshev collocation to the reduced equations. Finite-thickness jets
are treated separately through the steady closure condition. The spectral
solution converges geometrically, reproduces published steady data, and is
checked by componentwise off-grid residuals, temporal-tolerance studies and
an independent upwind discretisation. For fixed $\Fr$, $\We$, $\theta_0$
and pressure coefficient, the finite-thickness model predicts a convergence
length that decreases with thickness while the terminal zero remains
transversal; this is a statement about a specified path in parameter space,
not a universal experimental thickness law. The linear and nonlinear forced
responses both show a pressure-coefficient maximum near $\St=0.093$, whereas
the convergence-length gain increases monotonically over the globally
admissible interval $0.005\leq\St\leq0.405$. A global positivity check also
changes the interpretation of the first finite-time event. For body-force
forcing with $A=0.5$ and $\St_g=0.5$, $S$ first reaches zero at an interior
point, $t_c\simeq10.7538$ and $\eta_c\simeq0.9509$, while the computational
endpoint remains transversal, $S(t_c,1)\simeq0.484$. At contact,
$S_\eta=0$ and $S_{\eta\eta}>0$, so the mean radius is tangent to the axis
with quadratic local behaviour. An independent finite-difference scheme
converges to the same first-contact time. After this interior contact, the
tracked endpoint is no longer the first physical closure point, and the
single-interval annular-jet representation is no longer globally admissible.
\end{abstract}

\maketitle

\section{INTRODUCTION}
\label{sec:intro}

A liquid issuing from an annular nozzle forms a thin, approximately
axisymmetric sheet that may enclose a gas volume. If capillary forces are
sufficiently strong, the sheet contracts and reaches the symmetry axis at a
finite convergence length. This length, its dependence on the enclosed-gas
pressure, and its response to forcing are important observables in
applications ranging from liquid-curtain reactors to converging sheets in
pressure-swirl injectors.\cite{BairdDavidson1962b,Hoffman1980,Sivakumar1997,Sivakumar2002}
The capillary instability of annular jets and the liquid shells produced by
it have also been studied experimentally and theoretically.\cite{Kendall1986,LeeWang1986}
Resolved two-phase calculations can describe near-nozzle recirculation and
Kelvin--Helmholtz roll-up,\cite{SiamasJiang2009} phenomena outside the scope
of a one-dimensional theory. Such reduced models nevertheless remain useful
for the global sheet shape, enclosed volume and moving closure point.

The model used here is based on the integral formulation of
Ramos,\cite{Ramos1992ZAMM} obtained by averaging across a slender annular
sheet. Unsteady versions have been applied to mass loading, nozzle-flow
fluctuations, fluctuating body forces and gas absorption.\cite{RamosFalgueras1991,Ramos1992AMM,RamosFalgueras1992,Ramos1995,RamosIJNMF1995,RamosIJNMHFF1992,Ramos1993}
These calculations map the time-dependent interval onto a fixed unit
interval. The downstream closure point is not prescribed independently; it
follows from the condition that the inner radius vanish at the free end. In
the historical algorithms this algebraic closure relation is differentiated
in time, and the resulting equation for the convergence length is integrated
together with the field equations using extrapolated terminal
values.\cite{Ramos1993}

This treatment raises two distinct numerical questions. First, at the
continuous level the differentiated closure relation is equivalent to the
original one as long as closure is satisfied. In a discrete calculation,
however, integrating only the differentiated relation can allow drift away
from the closure condition unless the algebraic relation is also enforced or
stabilised. Second, satisfying the closure relation at the mapped endpoint
does not guarantee that this endpoint remains the \emph{first} point at
which the sheet reaches the axis. The radius must also remain positive over
the whole upstream interval. Either error can change the computed
convergence length or the interpretation of a periodic response.

For the membrane limit $\beta=0$, we remove the first difficulty by writing
$R=(1-\eta)S$. This representation satisfies $R(t,1)=0$ identically, so the
endpoint radius does not have to be obtained by extrapolation or restored by
projection. For finite thickness the physical closure condition is
$2R^2-\beta m=0$. An analogous algebraic representation can be written in
terms of the inner radius, as shown in Sec.~\ref{sec:model}; nevertheless,
all unsteady calculations reported here use the membrane equations. The
finite-thickness comparison in Sec.~\ref{sec:thickness} is restricted to
steady solutions, with the closure position located from the algebraic
condition.

A second issue is capillary criticality. The steady equations change type
when the liquid speed equals the sinuous-capillary wave speed. Ramos
identified the corresponding regularity condition for annular
jets,\cite{Ramos1996} which is closely related to the critical conditions
used for planar liquid sheets and bells.\cite{Brunet2004,LhuissierVillermaux2012,Chiatto2022}
In the subcritical regime one characteristic travels upstream and the usual
set of inlet conditions may overdetermine the problem, as discussed for
planar sheets.\cite{Benilov2019,Weinstein2019,Benilov2021,DellaPia2023}
We restrict the calculations below to the supercritical regime, for which
all characteristics enter from the nozzle and the stated inlet conditions
are appropriate. The subcritical boundary-value problem requires a different
specification and is outside the present scope. We nevertheless verify the
critical compatibility condition as an independent check on the steady
formulation.

The study has three aims. First, we formulate the membrane problem so that
the endpoint closure is satisfied identically and make explicit the
semidiscrete differential-algebraic structure that leads to the usual
evolution equation for the convergence length. Second, we verify the
numerical solution with tests that do not simply repeat the collocation
equations: componentwise residuals at off-grid points, decay of Chebyshev
coefficients, temporal-tolerance convergence, and a finite-difference
calculation based on different variables and stencils. Third, we use the
verified formulation to revisit three predictions of the model: the effect
of finite thickness on the steady convergence length, the response to
periodic forcing, and the first finite-time loss of physical admissibility.
The last result is changed by the global positivity check: before the mapped
endpoint loses transversality, the computed radius can reach the axis at an
interior point. At that time the tracked endpoint is still a simple zero,
but it is no longer the first closure station.

Chebyshev discretisations are well established for linear stability and
global eigenvalue problems of sheets and jets,\cite{DellaPia2020,Chiatto2022}
whereas the present calculation is nonlinear and has a moving boundary.
Section~\ref{sec:model} presents the model and the endpoint closure;
Sec.~\ref{sec:method} gives the membrane reduction and numerical
diagnostics; Sec.~\ref{sec:verify} verifies the steady and unsteady solvers;
Sec.~\ref{sec:thickness} treats the steady finite-thickness problem;
Sec.~\ref{sec:response} studies the forced response; and
Sec.~\ref{sec:transversality} identifies the first loss of global
admissibility. Section~\ref{sec:conclusions} concludes the paper.

\section{MODEL AND THE STRUCTURE OF THE TIP CONSTRAINT}
\label{sec:model}

\subsection{Governing equations and notation}

We use the one-dimensional asymptotic equations for an inviscid,
incompressible, axisymmetric annular liquid jet obtained by averaging the
Navier--Stokes equations across the
sheet.\cite{Ramos1992ZAMM,Ramos1993} Dimensional quantities are starred.
Let $z^\ast$ be the axial coordinate measured from the nozzle exit,
$t^\ast$ time, $R^\ast$ the mean radius of the sheet, $b^\ast$ its
thickness, $R_i^\ast=R^\ast-b^\ast/2$ and $R_e^\ast=R^\ast+b^\ast/2$ the
inner and outer radii, respectively, $u^\ast$ the axial velocity, $\bar v^\ast$ the
radial velocity averaged across the sheet, and
$m^\ast=\rho^\ast\,R^\ast\,b^\ast$ the liquid mass per unit axial length
and per radian, with $\rho^\ast$ the liquid density. Let $\sigma$ be the
surface tension coefficient, $g$ the gravitational acceleration, and
$p_i^\ast$ and $p_e^\ast$ the pressures of the gas enclosed by and
surrounding the sheet. The subscript $0$ denotes conditions at the nozzle
exit, and $\theta_0$ is the angle between the symmetry axis and the
velocity vector there.

Variables are made dimensionless with $R_0^\ast$ for lengths, $u_0^\ast$
for velocities, $R_0^\ast/u_0^\ast$ for time and $m_0^\ast$ for the mass
per unit length,
\begin{equation}
  z=\frac{z^\ast}{R_0^\ast},\quad
  t=\frac{t^\ast\,u_0^\ast}{R_0^\ast},\quad
  R=\frac{R^\ast}{R_0^\ast},\quad
  u=\frac{u^\ast}{u_0^\ast},\quad
  \bar v=\frac{\bar v^\ast}{u_0^\ast},\quad
  m=\frac{m^\ast}{m_0^\ast},
\end{equation}
so that the governing equations are
\begin{align}
  \frac{\partial m}{\partial t}
    + \frac{\partial}{\partial z}\,(m\,u) &= 0, \label{eq:mass}\\
  \frac{\partial}{\partial t}\,(m\,u)
    + \frac{\partial}{\partial z}\,(m\,u\,u)
    &= \frac{m}{\Fr}
      + \frac{1}{\We}\left(\frac{\partial J}{\partial z}
        - \Cpn\,R\,\frac{\partial R}{\partial z}\right), \label{eq:axial}\\
  \frac{\partial}{\partial t}\,(m\,\bar v)
    + \frac{\partial}{\partial z}\,(m\,u\,\bar v)
    &= \frac{1}{\We}\left(\Cpn\,R
        - \frac{\partial J/\partial z}{\partial R/\partial z}\right),
      \label{eq:radial}\\
  \bar v &= \frac{\partial R}{\partial t}
      + u\,\frac{\partial R}{\partial z}, \label{eq:kinematic}
\end{align}
where
\begin{equation}
  J = \frac{R}{\left[1+(\partial R/\partial z)^2\right]^{1/2}}
\end{equation}
is the geometric capillary term. Equation~(\ref{eq:mass}) expresses mass
conservation, Eqs.~(\ref{eq:axial}) and (\ref{eq:radial}) are the axial and
radial momentum balances, respectively, and Eq.~(\ref{eq:kinematic}) is the
kinematic condition at the mean radius. The dimensionless groups are the Froude
number, the Weber number and the pressure coefficient,
\begin{equation}
  \Fr = \frac{u_0^{\ast2}}{g\,R_0^\ast},\qquad
  \We = \frac{m_0^\ast\,u_0^{\ast2}}{2\,\sigma\,R_0^\ast},\qquad
  \Cpn = \Cpmax\left(\frac{p_i^\ast}{p_e^\ast}-1\right),\qquad
  \Cpmax = \frac{p_e^\ast\,R_0^\ast}{2\,\sigma}.
  \label{eq:groups}
\end{equation}
Thus $\Fr$ measures inertia relative to gravity and $\We$ inertia
relative to surface tension. The coefficient $\Cpn=(p_i^\ast-p_e^\ast)
R_0^\ast/(2\sigma)$ is the pressure difference across the sheet normalised
by the capillary-pressure scale $2\sigma/R_0^\ast$, while $\Cpmax$ is the
external gas pressure based on the same capillary scale. The thickness enters
through
\begin{equation}
  b = \beta\,\frac{m}{R},\qquad
  \beta = \frac{b_0^\ast}{R_0^\ast},
  \label{eq:thickness}
\end{equation}
$\beta$ being the thickness-to-radius ratio at the nozzle; $\beta=0$
defines an annular liquid \emph{membrane}, a sheet of vanishing
thickness. The nozzle conditions are
\begin{equation}
  m(t,0)=1,\quad u(t,0)=u_0(t),\quad
  \bar v(t,0)=u_0(t)\,\tan\theta_0,\quad R(t,0)=1 .
  \label{eq:nozzle}
\end{equation}
The gas enclosed by the sheet is taken to be ideal and isothermal and its
mass constant, so that
\begin{equation}
  \Cpn(t)=\Cpmax\left(\frac{p_i^\ast(0)}{p_e^\ast}\,
    \frac{V(0)}{V(t)}-1\right),\qquad
  V(t)=\int_0^{L(t)} R_i^2\,\dd z ,
  \label{eq:pressure}
\end{equation}
where $V$ is the enclosed volume divided by $\pi$ and $L$ is the
convergence length. Equation~(\ref{eq:pressure}) makes $\Cpn$ a
\emph{global} functional of the interface, which is the essential
difference from the ambient-pressure coupling of planar
curtains,\cite{Finnicum1993,Torsey2021,DellaPia2020,Chiatto2022} and it
is the coupling whose importance Sivakumar and Raghunandan established
experimentally.\cite{Sivakumar2002}

Note that $\beta$ appears neither in
Eqs.~(\ref{eq:mass})--(\ref{eq:kinematic}) nor in
Eq.~(\ref{eq:nozzle}), so that the thickness enters the problem only through the
tip relation and through the enclosed volume. This is used in
Sec.~\ref{sec:thickness}.

\subsection{Criticality of the steady system}
\label{sec:criticality}

Write $s=\partial R/\partial z$ and $q=1+s^2$. Using $m\,u=1$, which
follows from Eqs.~(\ref{eq:mass}) and (\ref{eq:nozzle}) in the steady
state with $u_0=1$, the steady form of
Eqs.~(\ref{eq:mass})--(\ref{eq:kinematic}) reduces to
\begin{equation}
  \begin{pmatrix} q & s\,u \\ \We\,s & \We\,u - R\,q^{-3/2}\end{pmatrix}
  \begin{pmatrix} \dd u/\dd z \\ \dd s/\dd z\end{pmatrix}
  =\begin{pmatrix} 1/(\Fr\,u) \\ \Cpn\,R - q^{-1/2}\end{pmatrix},
  \label{eq:steadysys}
\end{equation}
whose determinant is $\We\,u-J$. At the nozzle it vanishes when
$\We=\cos\theta_0$. This combination is already present in
Ref.~\onlinecite{Ramos1992ZAMM}, where the denominator $E-BD$ of its
Eqs.~(127)--(132) satisfies $\We\,(E-BD)=\We\,u-J$ identically. It is
the annular counterpart of the transcritical condition of planar sheets,
at which the liquid speed equals that of sinuous capillary
waves.\cite{Brunet2004,LhuissierVillermaux2012,Girfoglio2017} Ramos
obtained the same condition for the annular jet in the equivalent form
$\cos\theta=\We\,u/R$ and associated it with the critical Weber number at
which Baird and Davidson reported a slope discontinuity, an indeterminate
curvature and oscillations of the jet.\cite{Ramos1996}

At criticality the coefficient matrix is singular. A regular solution can
pass through the critical point only if the right-hand side lies in the
range of that matrix, which requires
\begin{equation}
  \Cpn = \Cpn^{*} \equiv
  \frac{\Fr\,R + \We^2\,s}{\Fr\,R^2\,\sqrt{1+s^2}} ,
  \label{eq:compat}
\end{equation}
equal to $\cos\theta_0+\cos^2\theta_0\,\sin\theta_0/\Fr$ at the nozzle
and to unity for $\theta_0=0$. In the small-slope limit
Eq.~(\ref{eq:compat}) reduces to $\Cpn=1+\We\,\tan\theta_0/\Fr$, which is
Eq.~(94) of Ref.~\onlinecite{Ramos1992ZAMM}, whose conclusions also state
the corresponding cylindrical branch. The same requirement, expressed as equality of the ranks of the
coefficient matrix and the augmented matrix, appears as Eq.~(35) of
Ref.~\onlinecite{Ramos1996}. The correspondence between the two notations is
as follows: unit volumetric flow rate, a Weber number twice the present one,
a pressure difference with the opposite sign and a different normalisation,
and $s=\tan\theta$. With these substitutions, Eq.~(35) of
Ref.~\onlinecite{Ramos1996} is equivalent to Eq.~(\ref{eq:compat}); its two
signs distinguish downward and upward jets. Neither
the critical surface nor the compatibility condition is therefore new
here; what the present formulation adds is that both are available in
closed form for arbitrary slope and are used systematically. Criticality therefore means
``singular coefficient matrix, check compatibility'', not ``no steady
state exists''. The role played here by Eq.~(\ref{eq:compat}) is the same
as that of the regularity condition that Chiatto and Della
Pia\cite{Chiatto2022} impose at the critical station of a subcritical
curtain in place of the second inlet boundary condition.

For completeness, applying the same elimination to the streamline
formulation of Eqs.~(B.14)--(B.17) in
Ref.~\onlinecite{Ramos1992ZAMM} gives the determinant $\We u-J/q$. Written
in terms of the streamline speed $V=u/\cos\theta$, its nozzle condition is
$\We V=\cos^2\theta_0$, which is the form used in
Ref.~\onlinecite{Ramos1997}. The apparently different factors
$\cos\theta_0$ and $\cos^2\theta_0$ therefore follow from the velocity
convention and the factor $(1+s^2)^{-1}$ that distinguishes the two
formulations; Appendix~B of Ref.~\onlinecite{Ramos1992ZAMM} discusses this
relation and estimates its effect on the convergence length to be at most
about fifteen per cent. The calculations below are restricted to the
supercritical regime, where the four characteristics enter through the
nozzle and the inlet conditions used here are appropriate.

\subsection{The tip relation as an algebraic closure condition}
\label{sec:constraint}

Map the time-dependent domain onto $[0,1]$ by $\eta=z/L(t)$, $\tau=t$.
The convergence point is the station at which the inner radius vanishes.
Using Eq.~(\ref{eq:thickness}), this physical closure condition becomes
\begin{equation}
  g\left(m(\tau,1),R(\tau,1)\right)\equiv
  2R^2-\beta m=0 \quad\text{at }\eta=1.
  \label{eq:constraint}
\end{equation}
For a membrane this reduces to $R(\tau,1)=0$. Physically,
Eq.~(\ref{eq:constraint}) is a consequence of closure; after the moving
domain has been mapped to $[0,1]$, it acts mathematically as an algebraic
constraint on the endpoint variables. It is Eq.~(76) of
Ref.~\onlinecite{Ramos1993}. Differentiation along the moving endpoint
gives
\begin{equation}
  \frac{\dd L}{\dd t}=
  \left.\frac{4R\left(uR_\eta-\bar vL\right)
        -\beta\,\partial_\eta(mu)}
       {4RR_\eta-\beta m_\eta}\right|_{\eta=1},
  \label{eq:Ldot}
\end{equation}
which is Eq.~(108) of Ref.~\onlinecite{Ramos1993}.

The differential-algebraic interpretation is most transparent if the endpoint
speed $w=\dot L$, rather than $L$ itself, is introduced as the algebraic
variable. After spatial semidiscretisation let $y$ collect the field degrees
of freedom together with the differential state $L$. The constrained system
has the semiexplicit form
\begin{equation}
   \dot y=F(\tau,y,w),\qquad 0=g(y),\qquad F_L=w.
   \label{eq:dae}
\end{equation}
One differentiation of the algebraic relation gives
$0=g_yF(\tau,y,w)$. In the membrane case the terminal kinematic equation
makes this relation
\begin{equation}
  0=\bar v(1)-\frac{u(1)-w}{L}R_\eta(1),
  \qquad
  \frac{\partial}{\partial w}(g_yF)=\frac{R_\eta(1)}{L}.
  \label{eq:dae1}
\end{equation}
Hence, while the endpoint zero is transversal, $R_\eta(1)\ne0$, the first
differentiation determines $w$ and a second differentiation determines
$\dot w$. This is the regular semiexplicit Hessenberg index-two structure.
The same statement for $\beta>0$ replaces $R_\eta(1)$ by the derivative of
the finite-thickness closure function along the endpoint.

Equation~(\ref{eq:Ldot}) is therefore a legitimate continuous index
reduction; the numerical issue is more specific. If the differentiated
relation is marched without simultaneously imposing $g=0$, truncation and
extrapolation errors can generate a component normal to the constraint
manifold. Projection or constraint stabilisation would control such drift.
The historical domain-adaptive implementations used the differentiated
length equation as the terminal evolution equation and did not additionally
project the numerical state onto Eq.~(\ref{eq:constraint}); our criticism
below is restricted to that discrete choice, not to index reduction in
itself. In the steady problem the differentiated equation carries no
independent information about $L$ once the time derivatives vanish, whereas
the original algebraic closure still selects the endpoint.

For the membrane problem, the most direct absorption is obtained by setting
\begin{equation}
  R(\tau,\eta)=(1-\eta)S(\tau,\eta).
  \label{eq:absorb}
\end{equation}
Then $R(\tau,1)=0$ holds identically for every $S$. For completeness, the
finite-thickness closure can also be absorbed algebraically. Define the inner
radius
\begin{equation}
 R_i=R-\frac{\beta m}{2R}=(1-\eta)\Sigma,
 \qquad
 R=\frac{R_i+\sqrt{R_i^2+2\beta m}}{2}.
 \label{eq:absorb-beta}
\end{equation}
Since $2R^2-\beta m=2RR_i$, Eq.~(\ref{eq:constraint}) is then satisfied
identically at $\eta=1$ for any $\beta\ge0$, and Eq.~(\ref{eq:absorb-beta})
reduces to Eq.~(\ref{eq:absorb}) when $\beta=0$. We give this transformation
to delimit the scope of the constraint argument; all unsteady spectral
calculations reported below use the membrane system $\beta=0$, while the
finite-thickness results of Sec.~\ref{sec:thickness} are steady closure-event
calculations. The zero at the tracked membrane endpoint is locally
transversal when
\begin{equation}
  S(\tau,1)=-L\,\frac{\partial R}{\partial z}\bigg|_{z=L}>0,
  \label{eq:transversality}
\end{equation}
and its speed is
\begin{equation}
  \frac{\dd L}{\dd t}=u(\tau,1)
       +\frac{L\bar v(\tau,1)}{S(\tau,1)}.
  \label{eq:LdotS}
\end{equation}
Local transversality is necessary but not sufficient for $L(t)$ to be the
first physical closure. Because $R=(1-\eta)S$, a single annular sheet on the
mapped interval additionally requires
\begin{equation}
   S(\tau,\eta)>0\qquad (0\leq\eta<1).
   \label{eq:globaladmiss}
\end{equation}
We call Eq.~(\ref{eq:globaladmiss}) global admissibility. The first failure
of either Eq.~(\ref{eq:transversality}) or Eq.~(\ref{eq:globaladmiss}) is the
first point at which the scalar endpoint $L(t)$ can no longer be interpreted
as the unique first closure of the annular jet. Section~\ref{sec:transversality}
shows that, in the principal forced case, the global condition fails first.

\section{SPECTRAL DISCRETISATION}
\label{sec:method}

\subsection{Reduced system and the removable singularity}

Substituting Eq.~(\ref{eq:absorb}) into the equations in the mapped domain and writing
$c=(u-\eta\,\dd L/\dd t)/L$ for the transport velocity in the mapped
frame gives
\begin{align}
  \frac{\partial m}{\partial\tau}
    &= -c\,\frac{\partial m}{\partial\eta}
       - \frac{m}{L}\,\frac{\partial u}{\partial\eta}, \label{eq:sm}\\
  \frac{\partial S}{\partial\tau}
    &= \frac{\bar v + c\,S}{1-\eta}
       - c\,\frac{\partial S}{\partial\eta}, \label{eq:sS}\\
  \frac{\partial u}{\partial\tau}
    &= -c\,\frac{\partial u}{\partial\eta} + \frac{1}{\Fr}
       + \frac{J_\eta-\Cpn\,R\,R_\eta}{m\,\We\,L}, \label{eq:su}\\
  \frac{\partial\bar v}{\partial\tau}
    &= -c\,\frac{\partial\bar v}{\partial\eta}
       + \frac{\Cpn\,R - J_\eta/R_\eta}{m\,\We}. \label{eq:sv}
\end{align}
The source of Eq.~(\ref{eq:sS}) is singular at $\eta=1$, but removably
so. With $G=\bar v+cS$, Eq.~(\ref{eq:LdotS}) gives $G(1)=0$, and the
analytical endpoint value used in the code is
\begin{equation}
   \left.\frac{G}{1-\eta}\right|_{\eta=1}=-G_\eta(1),
   \qquad
   S_\tau(1)=-G_\eta(1)-c(1)S_\eta(1).
   \label{eq:tiplimit}
\end{equation}
Thus no numerical $0/0$ is evaluated at the last collocation node. The quotient $J_\eta/R_\eta$ of
Eq.~(\ref{eq:sv}) is never formed as written, because under forcing $R$ may
develop interior extrema, at which numerator and denominator vanish
together, so the equivalent regular identity
\begin{equation}
  \frac{J_\eta}{R_\eta}
  = \frac{1}{\sqrt{q}}
    - \frac{R\,R_{\eta\eta}}{L^2\,q^{3/2}},
  \qquad q = 1 + \left(\frac{R_\eta}{L}\right)^{2},
  \label{eq:JoverR}
\end{equation}
is used at every node and at every time step. It is regular throughout
$[0,1]$, in particular at $\eta=1$, where $R$ vanishes.

The unknowns $m$, $S$, $u$ and $\bar v$ are represented at the $N+1$
Chebyshev--Gauss--Lobatto nodes of $[0,1]$. Their four nozzle values are
specified algebraically by Eq.~(\ref{eq:nozzle}), so each field contributes
$N$ evolved nodal values. Together with the differential state $L$, the
semidiscrete membrane problem therefore contains $4N+1$ ordinary
differential equations.

\subsection{Enclosed volume}

The pressure coefficient (\ref{eq:pressure}) requires
$V=L\int_0^1R^2\,\dd\eta$. If $S_N$ is the degree-$N$ Chebyshev interpolant,
then $R_N=(1-\eta)S_N$ has degree at most $N+1$ and $R_N^2$ degree at most
$2N+2$. Clenshaw--Curtis quadrature on the $N+1$ collocation nodes is not
exact for that polynomial. We therefore evaluate the pressure feedback with
an $(N+2)$-point Gauss--Legendre rule, which is exact through degree
$2N+3$ and hence integrates $R_N^2$ exactly.

This corrects an overstatement in the original manuscript, which used
$N+1$ Gauss points and described $R_N^2$ as degree $2N$. The numerical
impact is negligible at the resolutions used here but the exactness claim
matters. For the steady state with $\Fr=10$, $\We=50$ and $\Cpn=0.5$, the
relative Clenshaw--Curtis/Gauss difference is $2.1\times10^{-8}$ for $N=8$,
$2.4\times10^{-13}$ for $N=16$, and on the order of round-off from $N=24$. Replacing the
quadrature changes the rightmost eigenvalue by at most $5\times10^{-8}$ in
the tested cases. For the principal body-force calculation the corrected
rule changes the first-contact time by less than the digits quoted below.

\subsection{Steady solver}

The steady problem is solved by Newton's method with two continuation
sequences, one in the parameters from a base state and one in spatial
resolution, with barycentric transfer between resolutions. We define the
criticality distance
$\delta_c=\We-\cos\theta_0$. If parameter continuation is performed first on
a coarse grid and the solution is refined only afterwards, the Newton
iteration ceases to converge for $\delta_c\lesssim2\times10^{-2}$ because a
short near-nozzle scale is then unresolved. Increasing the resolution during
the parameter continuation extends the calculation to $\delta_c=10^{-2}$
with twelve significant digits (Table~\ref{tab:critical}). For smaller
$\delta_c$ we integrate Eq.~(\ref{eq:steadysys}) directly with a
Dormand--Prince method and terminate the integration when the closure
condition is reached; this calculation requires no continuation.

\subsection{A posteriori residual verification}
\label{sec:offgrid}

A collocation solution satisfies the spatially semidiscrete equations at the
nodes by construction, so an on-grid residual is not an independent spatial
check. We evaluate instead the residual of the interpolating polynomial at
$2001$ uniformly spaced points in
$0.05\leq\eta\leq1-10^{-6}$, none of which is used to impose an equation.
The lower cut removes the nozzle neighbourhood, where the four boundary
values are imposed algebraically rather than collocated. Consequently this
diagnostic is explicitly an \emph{interior spatial residual}; temporal error
is assessed separately by tolerance and integrator studies.

For each component $q\in\{m,S,u,\bar v\}$ we report
\begin{equation}
 r_q=\|q_\tau-F_q\|_\infty,
 \qquad
 \widehat r_q=\frac{r_q}
 {\max(\|q_\tau\|_\infty,\|F_q\|_\infty)},
 \label{eq:resnorm}
\end{equation}
where both norms are evaluated on the off-grid window. First and second spatial derivatives are obtained by differentiating the
$N$th-degree interpolant and evaluating the resulting polynomials on the fine
grid. The time derivatives are evaluated at the collocation nodes and
interpolated onto the same fine grid. We also audit the algebraic
boundary data directly: the four nozzle conditions, $R(1)=0$, and the
vanishing numerator in the removable endpoint quotient. These residuals are
on the order of round-off in the calculations reported below.

Spectral resolution is monitored independently. If $\widehat q_k$ are the
Chebyshev coefficients of a field, the tail diagnostic is
\begin{equation}
 T_q=\frac{\max(|\widehat q_{N-1}|,|\widehat q_N|)}
 {\max_{0\leq k\leq N}|\widehat q_k|}.
 \label{eq:chebtail}
\end{equation}
A decreasing $T_q$ under refinement is evidence that the represented field
remains spectrally resolved. Neither Eq.~(\ref{eq:resnorm}) nor
Eq.~(\ref{eq:chebtail}) is presented as a complete verification by itself;
they are used together with time-tolerance convergence and the independent
scheme of Sec.~\ref{sec:uw3}.

\subsection{An independent scheme}
\label{sec:uw3}

For cross-validation we use a method of lines that differs from the
spectral discretisation in the three respects that matter. It integrates
$R$ directly rather than $S$, so that Eq.~(\ref{eq:constraint}) is only
derived and not absorbed; it uses a third-order upwind-biased finite
difference on a uniform mesh with no global basis; and in the $R$
variables the quotient
$J_\eta/R_\eta=q^{-1/2}-R\,R_{\eta\eta}/(L^2\,q^{3/2})$ is regular at
$\eta=1$ because $R$ vanishes there, so the removable singularity of
Eq.~(\ref{eq:sS}) does not arise. Its observed order of convergence on the exact steady
state is two, the residuals being $1.674\times10^{-4}$,
$4.578\times10^{-5}$, $1.199\times10^{-5}$, $3.069\times10^{-6}$ and
$7.764\times10^{-7}$ for $M=40$, $80$, $160$, $320$ and $640$ mesh cells,
giving observed orders $1.87$, $1.93$, $1.97$ and $1.98$, limited by the
one-sided stencils at the nozzle. We note in passing that seeding
$\bar v=u\,R_\eta/L$ with the finite-difference derivative on the
scheme's own mesh, rather than with the spectral derivative of the seed,
introduces an $O(h^2)$ error in $\bar v$ that the scheme then
differentiates, degrading the residual to $O(h)$ and making a
second-order method appear first-order.

This scheme also illustrates the drift mechanism described in
Sec.~\ref{sec:constraint}. The endpoint speed is computed from
Eq.~(\ref{eq:LdotS}) using a centred approximation to $R_\eta(1)$, whereas
the advective term uses an upwind stencil. The two discrete derivatives are
not identical, so $R_\tau(1)$ is not exactly zero. A small error in $R(1)$
then changes $R_\eta(1)$, which changes $\dot L$ and feeds back into the
subsequent evolution of $R$. This feedback is unstable if the closure
condition is not imposed explicitly. In our tests $R_\eta(1)$ changed from
$-1.70$ to $+2.4$ before the integration broke down. Enforcing
Eq.~(\ref{eq:constraint}) at the final node removes this drift. Thus the
direct-$R$ formulation requires explicit enforcement of the endpoint
closure, whereas the $S$ formulation incorporates it identically.

\section{VERIFICATION}
\label{sec:verify}
\label{sec:verification}

\subsection{Spectral convergence and published steady data}

\begin{figure}[t]
\includegraphics[width=\linewidth]{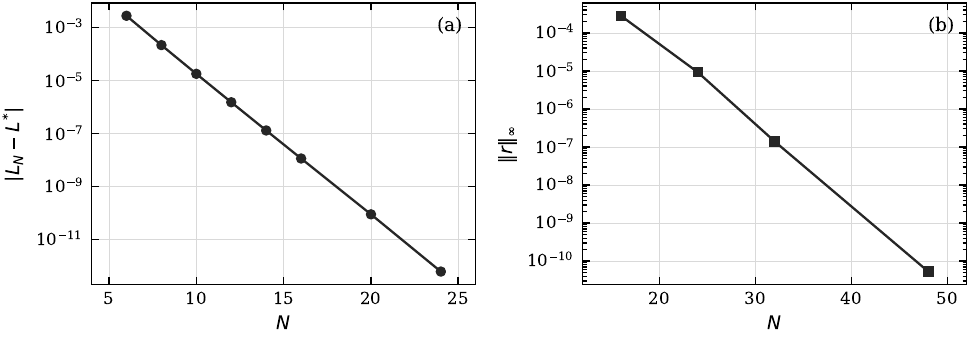}
\caption{Geometric convergence. (a) Error of the steady convergence
length against the resolution $N$ for $\Fr=10$, $\We=50$, $\theta_0=0$
and $\Cpn=0$; the reference is $N=96$. (b) Maximum off-grid residual of
the forced solution at the end of the integration for nozzle forcing with
$a_f=0.1$ and $\St=0.1$, for the same parameters.}
\label{fig:conv}
\end{figure}

Table~\ref{tab:conv} and Fig.~\ref{fig:conv}(a) show the convergence of
the steady convergence length at the base parameter set. The error falls
by a factor of about twelve per two added modes until it reaches
$6\times10^{-13}$ for $N=24$, where it saturates at the level of the
Newton tolerance.

\begin{table}[t]
\caption{Spectral convergence of the steady convergence length $L$ for
$\Fr=10$, $\We=50$, $\theta_0=0$ and $\Cpn=0$. The reference value is
$L^{*}=12.555846088897201$, obtained for $N=96$. The last column is the
ratio of successive errors. From $N=24$ the computed length agrees with
the reference in every digit carried in double precision, so the third
column no longer measures the error of the approximation but the floor of
the Newton solve; those entries are marked as saturated rather than
printed as zero, which would suggest an exactness the computation does not
establish.}
\label{tab:conv}
\begin{ruledtabular}
\begin{tabular}{rll r}
$N$ & $L$ & $|L-L^{*}|$ & factor \\
\colrule
 6 & 12.55864294603056 & $2.797\times10^{-3}$ & \\
 8 & 12.55606450542947 & $2.184\times10^{-4}$ & 12.81 \\
10 & 12.55586404578080 & $1.796\times10^{-5}$ & 12.16 \\
12 & 12.55584760703271 & $1.518\times10^{-6}$ & 11.83 \\
14 & 12.55584621955409 & $1.307\times10^{-7}$ & 11.62 \\
16 & 12.55584610028286 & $1.139\times10^{-8}$ & 11.48 \\
20 & 12.55584608898581 & $8.800\times10^{-11}$ & 129.4 \\
24 & 12.55584608889781 & saturated & \\
32 & 12.55584608889781 & saturated & \\
48 & 12.55584608889781 & saturated & \\
\end{tabular}
\end{ruledtabular}
\end{table}

Table~\ref{tab:ramos92} compares against the convergence lengths
tabulated by Ramos.\cite{Ramos1992AMM} Agreement is between
$1.2\times10^{-4}$ and $2.8\times10^{-3}$ in relative terms, consistent
with the first-order temporal and upwind spatial discretisation used
there. That table furnishes, in addition, an internal check of the same
kind as the one noted in Sec.~\ref{sec:thickness}. Three of its cases
differ only in $\Cpmax$, taking the values $1$, $2$ and $10$ at
$p_i^\ast(0)/p_e^\ast=1$, hence $\Cpn=0$ in all three, and all three are
reported with the same convergence length, confirming that $\Cpn$ is the
only pressure parameter of the steady problem.

\begin{table*}[t]
\caption{Comparison against the convergence lengths tabulated in Table~1
of Ref.~\onlinecite{Ramos1992AMM} for a membrane, computed here for
$N=32$. The second column is the critical margin and the last is the
Newton residual. The case for $\We=1$ of that table is treated separately
in Sec.~\ref{sec:nosteady}.}
\label{tab:ramos92}
\begin{ruledtabular}
\begin{tabular}{lrrrrl}
case & $\We-\cos\theta_0$ & $L$ (present) & $L$ (Ref.~\onlinecite{Ramos1992AMM}) & rel. diff. & $|F|$\\
\colrule
$\Fr=10$, $\We=50$, $\theta_0=0$      & 49.000 & 12.55584609 & 12.5590 & $2.51\times10^{-4}$ & $1.0\times10^{-11}$\\
$\Fr=10^4$, $\We=50$, $\theta_0=0$    & 49.000 &  9.88608175 &  9.8652 & $2.12\times10^{-3}$ & $1.1\times10^{-11}$\\
$\Fr=10$, $\We=100$, $\theta_0=0$     & 99.000 & 19.03734022 & 19.0452 & $4.13\times10^{-4}$ & $4.3\times10^{-12}$\\
$\Fr=10$, $\We=50$, $\theta_0=-15^{\circ}$ & 49.034 &  3.81309891 &  3.8236 & $2.75\times10^{-3}$ & $1.7\times10^{-10}$\\
$\Fr=10$, $\We=50$, $\theta_0=+15^{\circ}$ & 49.034 & 37.35123761 & 37.3466 & $1.24\times10^{-4}$ & $8.5\times10^{-13}$\\
$\Fr=10$, $\We=50$, $p_i^\ast(0)/p_e^\ast=0.5$  & 49.000 & 10.27485688 & 10.2816 & $6.56\times10^{-4}$ & $2.1\times10^{-11}$\\
\end{tabular}
\end{ruledtabular}
\end{table*}

\subsection{A case with no steady solution}
\label{sec:nosteady}

The case for $\Fr=10$, $\We=1$, $\theta_0=0$, $\Cpmax=1$ and
$p_i^\ast(0)/p_e^\ast=1$ deserves separate comment. It is reported with a
convergence length of $0.3262$ and used as the initial condition of an
unsteady computation, whereas the model admits no steady solution for these
parameters.

For this case the criticality distance $\delta_c=\We-\cos\theta_0$ is zero.
At the nozzle $s=0$ and $u=R=1$, so the
coefficient matrix of Eq.~(\ref{eq:steadysys}) becomes
$\mathrm{diag}(1,0)$ while its right-hand side is
$(1/\Fr,\,-1)^{\mathsf T}$; the second component is not in the range, and
by Eq.~(\ref{eq:compat}) compatibility would require $\Cpn=1$ rather than
the $\Cpn=0$ of that case. For $\We=1+\delta$ with $\delta>0$ the system
is regular but
\begin{equation}
  \frac{\dd s}{\dd z}\bigg|_{z=0}=-\frac{1}{\delta},
  \label{eq:slopeblowup}
\end{equation}
so the slope of the interface at the nozzle diverges as the margin
vanishes and the solution develops a layer of width $O(\delta)$ there.
This is the mechanism behind the stalling of the parameter continuation
reported in Sec.~\ref{sec:method} and quantified in
Table~\ref{tab:critical}.

The convergence length itself does not vanish in that limit. A
Dormand--Prince march gives $L=0.10073$, $0.10728$, $0.15244$ and
$0.28607$ at $\delta=10^{-4}$, $10^{-3}$, $10^{-2}$ and
$5\times10^{-2}$, and $L\to0.099752$ as $\delta\to0^{+}$. The value
$0.3262$ corresponds instead to a margin $\delta=0.0650$, that is to
$\We=1.0650$. A first-order scheme on a uniform grid cannot resolve a
nozzle layer whose width tends to zero, and returns a finite and
superficially plausible convergence length for a parameter set at which
the model has none. Note also that Eq.~(\ref{eq:slopeblowup}) violates
the small-slope assumption under which the asymptotic model was derived,
so that neither value should be regarded as physically meaningful; the
point is that the discrete problem gives no indication of the difficulty.

This conclusion is not in conflict with the later literature but
anticipated by it. Ramos\cite{Ramos1996} reports that numerical
experiments in which the mass flow rate at the nozzle was reduced towards
the critical value produced no annular liquid jet at unit Weber number
with a vanishing nozzle angle, and attributes this to the singularity of
the coefficient matrix at the nozzle exit. What Eqs.~(\ref{eq:compat})
and (\ref{eq:slopeblowup}) add is the reason, namely the failure of
compatibility and the rate at which the nozzle layer collapses, and a
quantitative account of what a scheme that cannot resolve that layer
returns instead.

\begin{table}[t]
\caption{Approach to criticality for $\Fr=10$, $\theta_0=0$ and
$\Cpn=0$, computed by collocation for $N=64$ and by a Dormand--Prince
integration of Eq.~(\ref{eq:steadysys}). The criticality distance is
$\delta_c=\We-\cos\theta_0$. The collocation solver does not converge for
the smallest value shown.}
\label{tab:critical}
\begin{ruledtabular}
\begin{tabular}{rrrrl}
$\We$ & margin & $L$ (collocation) & $L$ (march) & rel. diff.\\
\colrule
1.5000 & 0.5000 & 0.9987763984 & 0.9987763984 & $1.2\times10^{-13}$\\
1.2000 & 0.2000 & 0.5988803429 & 0.5988803429 & $1.1\times10^{-13}$\\
1.1000 & 0.1000 & 0.4089802752 & 0.4089802752 & $1.4\times10^{-13}$\\
1.0500 & 0.0500 & 0.2860713896 & 0.2860713896 & $1.1\times10^{-13}$\\
1.0200 & 0.0200 & 0.1916928423 & 0.1916928423 & $3.3\times10^{-13}$\\
1.0100 & 0.0100 & 0.1524367441 & 0.1524367441 & $1.9\times10^{-12}$\\
1.0050 & 0.0050 & 0.1364630913 & 0.1294788741 & $5.4\times10^{-2}$\\
1.0010 & 0.0010 & \multicolumn{1}{c}{---} & 0.1072779141 & \\
\end{tabular}
\end{ruledtabular}
\end{table}

\subsection{Unsteady verification}

Table~\ref{tab:offgrid} reports the forced response at forcing amplitude
$a_f=0.1$ and $\St=0.1$, with
$\St=\omega^\ast R_0^\ast/(2\pi u_0^\ast)$. To avoid confusion with the
mapped transport velocity $c$ of Sec.~\ref{sec:method}, $a_f$ denotes only
the nozzle-forcing amplitude. The imposed velocity is
\begin{equation}
 u_0(\tau)=1+a_f E\!\left(\frac{\tau}{2T}\right)
 \sin(2\pi\St\tau),\qquad T=\St^{-1},
 \label{eq:nozzleforcing}
\end{equation}
where the initial phase is zero and
\begin{equation}
 E(x)=\begin{cases}
 0,&x\le0,\\
 \displaystyle\frac{e^{-1/x}}{e^{-1/x}+e^{-1/(1-x)}},&0<x<1,\\
 1,&x\ge1
 \end{cases}
 \label{eq:rampenvelope}
\end{equation}
is a $C^\infty$ two-cycle start-up envelope. The integration in
Table~\ref{tab:offgrid} lasts nine cycles and the mean, fundamental amplitude
and time lag are fitted over the final three. The mean convergence length is
stable to eight digits between $N=32$ and $N=48$, the implicit Radau
integrator reproduces the explicit one at the same resolution, and the
interior residual falls geometrically from $2.8\times10^{-4}$ to
$5.3\times10^{-11}$ [Fig.~\ref{fig:conv}(b)]. A global audit gives
$S>0$ throughout all five runs.

\begin{table}[t]
\caption{Forced response for $a_f=0.1$, $\St=0.1$, $\Fr=10$,
$\We=50$ and $\theta_0=0$, with the two-cycle envelope
(\ref{eq:rampenvelope}). Here $A_L$ is the fundamental amplitude of $L$;
$\Delta\tau_L$ is the dimensionless time delay of $L$ behind the imposed
nozzle velocity; and the last column is the absolute maximum interior
off-grid residual.}
\label{tab:offgrid}
\begin{ruledtabular}
\begin{tabular}{rlrrrl}
$N$ & integrator & $\langle L\rangle$ & $A_L/\langle L\rangle$ &
$\Delta\tau_L$ & $\max_q r_q$\\
\colrule
16 & DOP853 & 12.48787669 & 0.12659 & 7.4045 & $2.82\times10^{-4}$\\
24 & DOP853 & 12.48785957 & 0.12658 & 7.4045 & $9.26\times10^{-6}$\\
32 & DOP853 & 12.48785948 & 0.12658 & 7.4045 & $1.40\times10^{-7}$\\
48 & DOP853 & 12.48785948 & 0.12658 & 7.4045 & $5.34\times10^{-11}$\\
24 & Radau  & 12.48785957 & 0.12658 & 7.4045 & $9.26\times10^{-6}$\\
\end{tabular}
\end{ruledtabular}
\end{table}

The rightmost eigenvalue of the semidiscrete operator, with the pressure
coupling of Eq.~(\ref{eq:pressure}) retained in the linearisation, is
negative for every tested resolution. Thus the discrete steady state is
linearly stable throughout the tested range. The rightmost eigenvalue itself
is not converged: it varies between $-0.25$ and $-0.34$ for
$8\le N\le48$ because the identity of the rightmost mode changes. By
contrast, the least-damped oscillatory pair
$-0.31038+0.63253\,\mathrm{i}$, reproduced for $N=24$ and $N=32$, whose
period $9.9334$ governs the frequency response of
Sec.~\ref{sec:response}.

\section{JETS OF FINITE THICKNESS}
\label{sec:thickness}

The result of this section rests on a structural property of the model
that is worth stating explicitly, because it is what makes the argument
exact rather than numerical. In the non-dimensional formulation of
Ref.~\onlinecite{Ramos1993}, conservation of mass, the two momentum
balances and the kinematic condition, its Eqs.~(17)--(20), reproduced
here as Eqs.~(\ref{eq:mass})--(\ref{eq:kinematic}), involve only $m$,
$u$, $\bar v$, $R$ and $J$. The thickness appears in that paper solely in
its Eqs.~(22)--(24), which define $R_e$, $R_i$ and $b=\beta\,m/R$, and in
the tip relation, its Eq.~(76). The same is true of the perturbation
derivation of Ref.~\onlinecite{RamosIJNMF1995}, whose hydraulic
equations coincide term by term with
Eqs.~(\ref{eq:mass})--(\ref{eq:kinematic}). There are therefore no
$O(\beta)$ corrections to the momentum balances within this formulation.
The effect discussed below enters only through the closure condition and the
enclosed volume.

\begin{figure}[t]
\includegraphics[width=\linewidth]{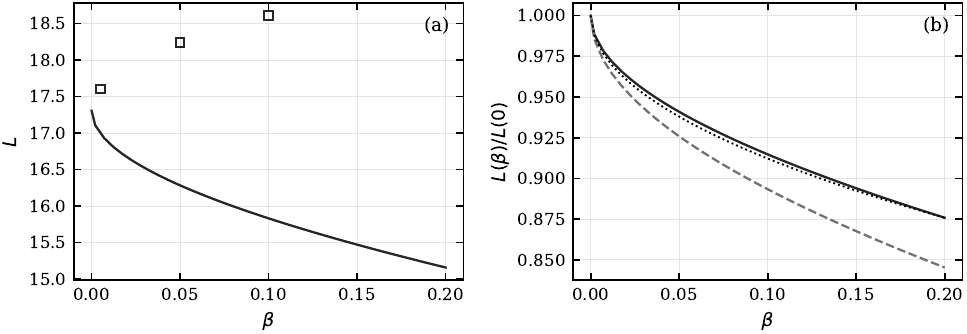}
\caption{Convergence length against the nozzle thickness-to-radius ratio.
(a) Solid line, the prediction of the asymptotic model for $\Fr=10$,
$\We=50$ and $\Cpn=0.5$; open squares, the three values tabulated in
Table~4 of Ref.~\onlinecite{Ramos1993} for the same parameters. The model
prediction decreases with $\beta$, whereas the tabulated values increase.
(b) The same length normalised by the membrane limit $L(0)$: solid line,
$\We=50$ and $\Cpn=0.5$;
dashed line, $\We=25$ and $\Cpn=0$; dotted line, the square-root law
$1-c\,\beta^{1/2}$ with the coefficient $c=4.377$ obtained from the
small-$\beta$ data of Sec.~\ref{sec:thickness}, and extended here to
$\beta=0.2$ to show where it ceases to describe the curve.}
\label{fig:thickness}
\end{figure}

\begin{figure}[t]
\includegraphics[width=\linewidth]{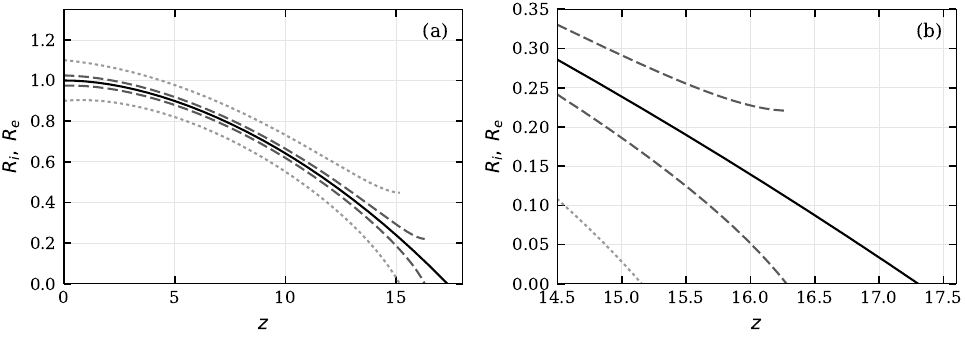}
\caption{Steady inner and outer interfaces for $\Fr=10$, $\We=50$,
$\theta_0=0$ and $\Cpn=0.5$, with $\beta=0$ (solid), $\beta=0.05$ (dashed)
and $\beta=0.2$ (dotted). (a) Whole jet. (b) Formal closure region. The mean
radius is the same curve in the three cases; only the station at which the
inner interface reaches the axis moves upstream as $\beta$ increases. For
$\beta>0$ the thin-sheet approximation is least reliable in this closure
region because the local thickness becomes comparable with the mean radius;
panel (b) is included to show where the model places its formal closure.}
\label{fig:profiles}
\end{figure}

Fix now $\Fr$, $\We$, $\theta_0$ and $\Cpn$. Because the thickness $b$
appears neither in Eqs.~(\ref{eq:mass})--(\ref{eq:kinematic}) nor in the
nozzle data (\ref{eq:nozzle}), the solution of Eq.~(\ref{eq:steadysys})
is independent of $\beta$, which enters only through
Eq.~(\ref{eq:constraint}). Writing the convergence length as the first
zero of $h(z)=R(z)^2-\beta\,m(z)/2$ and differentiating
$h(L(\beta),\beta)=0$ implicitly, with
$\partial h/\partial\beta=-m/2<0$, gives
\begin{equation}
  \frac{\dd L}{\dd\beta}=\frac{m(L)/2}{h'(L)}<0
  \label{eq:mono}
\end{equation}
whenever the zero is transversal, $h'(L)<0$. The convergence length is
therefore a strictly decreasing function of the thickness-to-radius ratio
at the nozzle wherever that hypothesis holds, \emph{when $\Fr$, $\We$,
$\theta_0$ and $\Cpn$ are held fixed}. This is a partial derivative in the
model parameter space. It is not the total derivative along every physical
experiment. For example, if the mass and temperature of the same enclosed
gas are held fixed while $\beta$ changes, the enclosed volume changes and
$\Cpn$ must change with it, so Eq.~(\ref{eq:mono}) alone does not determine
the sign of the resulting total derivative. Likewise, from
Eq.~(\ref{eq:groups}) the Weber number changes with nozzle thickness when
fluid properties, nozzle radius and dimensional velocity are fixed. A
fixed-$\We$ thickness sweep therefore represents a controlled mathematical
comparison in which the corresponding dimensional quantities are adjusted
to keep $\We$ constant; it should not be read as a universal experimental
thickness law. The transversality condition holds at every sampled point. We tested more
than forty combinations of $\We$ between $2$ and $200$ and $\theta_0$
between $-30^{\circ}$ and $+30^{\circ}$, for five values of $\Cpn$ and five
values of $\beta$. The largest value of $h'(L)$
encountered was $-2.6\times10^{-3}$ for $\We=200$, $\theta_0=0$,
$\Cpn=0.5$ and $\beta=0.005$. Transversality was thus retained throughout
the sampled set, with the margin smallest at large Weber number and small
thickness; we do not extrapolate that to a statement about the limit
$\beta\to0^{+}$. Moreover,
near the tip $R$ vanishes linearly, so the level $\sqrt{\beta\,m/2}$ that
defines the free end rises like $\beta^{1/2}$ and
$L(0)-L(\beta)=c\,\beta^{1/2}+o(\beta^{1/2})$ with $c>0$; hence
$\dd L/\dd\beta\to-\infty$ as $\beta\to0^{+}$ and the membrane is not a
regular limit of the jet of finite thickness. Table~\ref{tab:beta},
Fig.~\ref{fig:thickness} and Fig.~\ref{fig:profiles} verify both
statements.

The exponent is measured rather than assumed. Fitting
$\log[L(0)-L(\beta)]$ against $\log\beta$ by least squares over $100$
values logarithmically spaced in $10^{-4}\le\beta\le10^{-2}$ gives an
exponent of $0.5034$ with $c=4.516$ for $\We=50$ and $\Cpn=0.5$, and
$0.5029$ with $c=2.731$ for $\We=25$ and $\Cpn=0$, the fit reproducing the
data to better than $4\times10^{-3}$ in relative terms in both families.
The residual excess over $1/2$ is the neglected $o(\beta^{1/2})$ term and
decreases as the range shrinks. Taken decade by decade in the first family
the local exponent is $0.5073$, $0.5040$, $0.5022$ and $0.5012$ from the
highest decade to the lowest. Correspondingly, the ratio
$[L(0)-L(\beta)]\,\beta^{-1/2}$ is monotone and tends to $c$; a fit of the
form $c+k\,\beta^{1/2}$ over the twenty smallest values extrapolates to
$c=4.377$.

Two caveats are important. First, this is a statement about the
one-dimensional model, not a direct statement about the physical flow. At the closure point
the tip relation gives $b=2\,R$, so the sheet is locally as thick as its
own mean radius and the thin-sheet hypothesis under which
Eqs.~(\ref{eq:mass})--(\ref{eq:kinematic}) were derived is violated for
every $\beta>0$; the same objection is raised in
Ref.~\onlinecite{Ramos1992ZAMM}, which estimates the extent of the
invalid region near the convergence point and finds it to vanish only in
the membrane limit. The monotonicity above is therefore a property of the
equations as posed, and the region in which it is being applied is
precisely the region in which those equations are least trustworthy. The
second is that the argument assumes transversality, which
Sec.~\ref{sec:transversality} shows may fail.

\begin{table}[t]
\caption{Convergence length as a function of the nozzle
thickness-to-radius ratio for two parameter families with $\Fr=10$ and
$\theta_0=0$. The values of $(\We,\Cpn)$ are shown in the column headings.
For both families $L$ decreases with $\beta$, as required by
Eq.~(\ref{eq:mono}).}
\label{tab:beta}
\begin{ruledtabular}
\begin{tabular}{rrr}
$\beta$ & $L$ ($\We=50$, $\Cpn=0.5$) & $L$ ($\We=25$, $\Cpn=0$)\\
\colrule
0.000 & 17.304195 & 8.323897\\
0.005 & 16.990499 & 8.133657\\
0.010 & 16.858043 & 8.053575\\
0.050 & 16.281523 & 7.706632\\
0.100 & 15.829237 & 7.436184\\
0.200 & 15.155704 & 7.035977\\
\end{tabular}
\end{ruledtabular}
\end{table}

Table~\ref{tab:ramos93} compares against Table~4 of
Ref.~\onlinecite{Ramos1993}, which tabulates the enclosed gas mass, the
enclosed volume and the convergence length for eleven parameter sets at
$\beta=0.05$. Two internal consistency checks of that table are
satisfied. Two different pairs $(\Cpmax,\,p_i^\ast(0)/p_e^\ast)$ that
give the same $\Cpn=0.5$ produce the same convergence length, confirming
that $\Cpn$ is the only pressure parameter of the steady problem; and
the tabulated gas mass equals $1.5$ times the tabulated volume
throughout, as required by its Eq.~(58). The tabulated lengths are
nevertheless systematically about six per cent above the membrane-limit length
where the exact ones lie about six per cent below it, and, for $\We=50$,
they increase with $\beta$, a behaviour on which that paper explicitly
remarks by contrast with its $\We=25$ family. By Eq.~(\ref{eq:mono})
this cannot be a property of the equations.

\begin{table*}[t]
\caption{Comparison against Table~4 of Ref.~\onlinecite{Ramos1993} for
$\beta=0.05$ except where stated, with $\theta_0=0$ throughout;
$r$ denotes $p_i^\ast(0)/p_e^\ast$ and unlisted parameters are
$\Fr=10$ and $\We=50$. The last column is the ratio of the tabulated
value to the exact membrane-limit length.}
\label{tab:ramos93}
\begin{ruledtabular}
\begin{tabular}{lrrrrr}
case & $\Cpn$ & $L$ ($\beta=0$) & $L$ ($\beta$) & $L$ (Ref.~\onlinecite{Ramos1993}) & ratio\\
\colrule
$r=0.5$                 & $-0.50$ & 10.27486 &  9.48589 & 11.030 & 1.073\\
$r=1.0$                 &  $0.00$ & 12.55585 & 11.67013 & 13.388 & 1.066\\
$\Cpmax=1$, $r=1.05$    &  $0.05$ & 12.87031 & 11.97305 & 13.712 & 1.065\\
$\Cpmax=5$, $r=1.05$    &  $0.25$ & 14.39585 & 13.44805 & 15.278 & 1.061\\
$\Cpmax=10$, $r=1.05$   &  $0.50$ & 17.30420 & 16.28152 & 18.239 & 1.054\\
$\beta=0.005$           &  $0.50$ & 17.30420 & 16.99050 & 17.604 & 1.017\\
$\beta=0.1$             &  $0.50$ & 17.30420 & 15.82924 & 18.609 & 1.075\\
$\Fr=1000$              &  $0.50$ & 13.09480 & 12.13239 & 13.930 & 1.064\\
$\Fr=\infty$            &  $0.50$ & 13.03854 & 12.07776 & 13.872 & 1.064\\
$\We=5$                 &  $0.50$ &  4.25063 &  3.94389 &  4.511 & 1.061\\
$\We=75$                &  $0.50$ & 22.13138 & 20.87351 & 23.289 & 1.052\\
\end{tabular}
\end{ruledtabular}
\end{table*}

The sign conflict with Table~4 of Ref.~\onlinecite{Ramos1993} strongly
suggests that the free-end treatment is one possible source of the
discrepancy, because the historical calculation marches
Eq.~(\ref{eq:Ldot}) with extrapolated terminal values whereas the present
steady calculation locates Eq.~(\ref{eq:constraint}) directly. This
comparison does \emph{not} uniquely identify the cause: without reproducing
the complete historical implementation, differences in parameter
interpretation, transcription, or other numerical details cannot be
excluded. It is consistent with a free-end contribution that the same
historical scheme reproduces the membrane lengths of
Table~\ref{tab:ramos92} much more closely, and that the enclosed volume
tabulated for $\beta=0.05$ and $\Cpn=0.5$ is close to the membrane volume; these
observations are supporting evidence, not a causal proof.

Experimental convergence lengths for thin annular jets are reported by
Sivakumar and Raghunandan\cite{Sivakumar1997} over
$\beta\in[2.2\times10^{-3},\,7.0\times10^{-3}]$, a range in which
Eq.~(\ref{eq:mono}) predicts a variation of well under one per cent;
consistently, they find the measured length to collapse onto a single
curve against the mass flow rate independently of $\beta$. Their data lie
below the predictions of the one-dimensional model, which they attribute
in part to the different operational definitions of the convergence point
in experiment and in theory.

\section{FORCED RESPONSE}
\label{sec:response}

\subsection{Quasi-static gains}

The reference against which a frequency response must be normalised is
the quasi-static gain, and obtaining it requires care. Perturbing $u_0$
in a steady solver that normalises the mass flux to $m\,u=1$ is not the
operation performed by the unsteady boundary condition
(\ref{eq:nozzle}), which holds $m(t,0)=1$ and lets the nozzle mass flux
$F_0=m_0\,u_0$ vary. Repeating the elimination of
Sec.~\ref{sec:criticality} with general $F_0$ gives
\begin{equation}
  \frac{\dd s}{\dd z}\left[F_0\,u-\frac{J}{\We}\right]
  = \frac{q\,\left(\Cpn\,R-q^{-1/2}\right)}{\We}
    -\frac{F_0\,s}{u\,\Fr},
  \label{eq:generalF0}
\end{equation}
which reduces to Eq.~(\ref{eq:steadysys}) for $F_0=1$. The critical
condition becomes $\We F_0u=J$, or $\We u_0^2=\cos\theta_0$ at the nozzle,
so perturbing the nozzle velocity also changes the distance from
criticality. The relevant response is a closed-loop gain because a change in
$L$ changes $V$ and therefore $\Cpn$ through Eq.~(\ref{eq:pressure}). For
the base state $\Fr=10$, $\We=50$, $\theta_0=0$, $\Cpmax=1$ and
$\Cpn=0$, we obtain $\dd L/\dd u_0=10.571037$ when $\Cpn$ is artificially
held fixed at zero, $\dd L/\dd u_0=7.035783$ when the gas-pressure feedback
is included, and $\dd\Cpn/\dd u_0=-0.582084$. The corresponding normalised
closed-loop length gain is $0.560359$. The negative sign of the latter
derivative means that $\Cpn$ responds in antiphase: increasing the nozzle
velocity lengthens the jet, increases the enclosed volume and lowers
$\Cpn$.

The unsteady solver approaches these values as the forcing frequency is
reduced. For $\St=0.005$ and $a_f=0.02$, the measured length and pressure
gains are $0.56239$ and $0.58317$, differing from the independent
quasi-static values by $0.4\%$ and $0.2\%$, respectively. The two results are
obtained by different calculations: a Dormand--Prince integration of the
steady equations and a spectral time integration of the unsteady system.
Their agreement therefore provides a direct consistency check on the
unsteady implementation. We repeated this check for six additional base
states, including $\We=25$ and $100$, $\Fr=1000$, and pressure differences
of both signs. Across these tests the low-frequency gains agree with the
independent static derivatives to within $0.72\%$ for $L$ and $0.42\%$ for
$\Cpn$. This comparison is used only as a numerical verification of the
quasi-static limit; it is not intended as a universal statement about the
rate at which that limit is reached.

\subsection{Frequency response}

For the frequency-response calculations the nozzle velocity is given by
Eq.~(\ref{eq:nozzleforcing}) with $a_f=0.02$. For each value of $\St$, the
integration contains
\begin{equation}
 N_c=\max\left(8,\left\lceil70/T\right\rceil+5\right)
 \label{eq:ncycles}
\end{equation}
cycles. The first $N_c-3$ cycles are discarded. The mean and fundamental
amplitudes reported in Table~\ref{tab:freq} are obtained, exactly as in the
deposited table driver, by a least-squares fit of the final three cycles to a
constant plus sine and cosine at the forcing frequency. To quantify the
higher harmonics independently, the same interval is also refitted to a
constant plus sine/cosine pairs at the first four harmonics. A periodic
response is accepted only after the maximum relative difference between two
successive phase-aligned cycles of both $L$ and $\Cpn$ is below $10^{-8}$.
For $\St=0.1$, for example, those differences are $2.77\times10^{-13}$ and
$2.40\times10^{-12}$ for $N=32$. The four-harmonic diagnostic gives
$A_2/A_1=3.68\times10^{-2}$ for $L$ and $1.39\times10^{-2}$ for $\Cpn$;
the corresponding third-harmonic ratios are $2.09\times10^{-3}$ and
$3.27\times10^{-4}$, and the fourth-harmonic ratios are
$1.46\times10^{-4}$ and $1.16\times10^{-5}$. The simultaneous
four-harmonic refit changes the $L$ fundamental gain from $1.28286$ to
$1.28242$ ($0.035\%$) and the $\Cpn$ gain from $0.83110$ to $0.831108$
($0.001\%$), so the extraction choice has no effect on the conclusions.
Changing $N$ from 24 to 48 or the relative tolerance from $10^{-9}$ to
$10^{-11}$ changes the four-harmonic gains only beyond the digits relevant
here. Halving and doubling $a_f$ to $0.01$ and $0.04$ changes the
four-harmonic fundamental gain of $L$ by less than $0.25\%$ for $\St=0.1$,
while the higher harmonics scale as expected for the weakly nonlinear
response.

Table~\ref{tab:freq} reports the nonlinear gains divided by their
quasi-static values; both normalised gains therefore approach unity as
$\St\to0$. Over the interval for which a periodic solution remains globally
admissible, $0.005\le\St\le0.405$, the convergence-length gain increases
monotonically. This was checked on a dense sweep of 79 frequencies. The upper
limit is set by the solution itself: for $\St=0.410$ and $N=32$, an interior
contact occurs at $t_c=14.2716$ and $\eta_c\simeq0.9734$, while the endpoint
still satisfies $S(t_c,1)\simeq0.096$. For $\St=0.405$, no contact is found
during the converged periodic calculation and the minimum of $S$ remains
positive. Consequently, the present data neither show nor exclude a
resonance of the convergence length outside the admissible interval.

The pressure-coefficient response, by contrast, has a well-defined resonance
within this interval. The nonlinear dense sweep gives a maximum normalised
gain of $1.441$ for $\St\simeq0.0925$. We checked this result independently
in the small-amplitude limit. Let $J$ denote the Jacobian of the closed-loop
semidiscrete system at the steady state and $B$ the derivative of its vector
field with respect to the imposed nozzle velocity. The linear transfer
function
\begin{equation}
   H(\omega)=C\,(\mathrm{i}\omega I-J)^{-1}B
   \label{eq:transfer}
\end{equation}
recovers the independently calculated quasi-static gains at $\omega=0$:
$0.56035917$ for $L/\langle L\rangle$ and $0.58208418$ for $\Cpn$. Its
pressure response peaks for $\St=0.093$ with normalised gain $1.4407$, while
its convergence-length gain remains monotone through $\St=0.405$. Thus the
pressure peak is a property of the linearised semidiscrete dynamics and not
of the start-up protocol or the harmonic extraction.

The least-damped complex pair,
$-0.31038+0.63253\,\mathrm{i}$, has natural frequency $0.7046$ and damping
ratio $\zeta=0.4405$. If this pair were represented by an isolated
second-order oscillator, its amplitude maximum would occur at
$\omega_n\sqrt{1-2\zeta^2}=0.551$. The full linear system peaks instead near
$\omega=2\pi(0.093)=0.584$ because the transfer function also contains the
remaining modes and their non-normal coupling. Equation~(\ref{eq:transfer})
therefore provides the quantitative comparison; the single-mode estimate is
included only to relate the resonance to the least-damped oscillatory part of
the spectrum.

\begin{table}[t]
\caption{Frequency response for $a_f=0.02$, $N=32$, $\Fr=10$, $\We=50$
and $\theta_0=0$, normalised by the closed-loop quasi-static gains. Here
$A_L$ and $A_{\Cpn}$ are the fundamental amplitudes of $L$ and $\Cpn$.
Among the tabulated frequencies the pressure gain is largest for $\St=0.100$;
a dense frequency sweep gives the actual maximum for $\St\simeq0.0925$.}
\label{tab:freq}
\begin{ruledtabular}
\begin{tabular}{rrrrr}
$\St$ & $A_L/(a_f\,\langle L\rangle)$ & normalised &
$A_{\Cpn}/a_f$ & normalised\\
\colrule
0.005 & 0.56239 & 1.004 & 0.58317 & 1.002\\
0.010 & 0.56849 & 1.015 & 0.58630 & 1.007\\
0.020 & 0.59290 & 1.058 & 0.59886 & 1.029\\
0.050 & 0.76507 & 1.365 & 0.68659 & 1.180\\
0.080 & 1.07586 & 1.920 & 0.81482 & 1.400\\
0.100 & 1.28286 & 2.289 & 0.83110 & 1.428\\
0.125 & 1.40440 & 2.506 & 0.69182 & 1.189\\
0.150 & 1.46789 & 2.620 & 0.50486 & 0.867\\
0.200 & 1.74909 & 3.121 & 0.29152 & 0.501\\
0.300 & 2.28612 & 4.080 & 0.17049 & 0.293\\
0.400 & 2.57704 & 4.599 & 0.10891 & 0.187\\
\end{tabular}
\end{ruledtabular}
\end{table}

Phase lags must be reported with care. A lag defined modulo the forcing
period is ambiguous, and the pressure coefficient is in antiphase in the
quasi-static limit, so its phase must be reduced by $\pi$ before a lag is
read off. Unwrapping in $\St$ from the quasi-static limit gives lags for
the convergence length of $6.11$ as $\St\to0$, rising to $7.37$ for
$\St=0.1$, and for the pressure coefficient of $3.38$ rising to $4.68$.
These are computed for $a_f=0.02$; the value $7.4045$ quoted in
Table~\ref{tab:offgrid} for the same Strouhal number differs from $7.37$
because it corresponds to the larger amplitude $a_f=0.1$, at which the
response is no longer strictly linear.
The pressure coefficient therefore responds ahead of the convergence
length at every frequency, which is the qualitative statement of
Refs.~\onlinecite{RamosFalgueras1992} and \onlinecite{Ramos1995} made
quantitative.

\section{FINITE-TIME LOSS OF GLOBAL ADMISSIBILITY}
\label{sec:transversality}

The membrane substitution enforces the mapped endpoint zero exactly, but it
does not guarantee that this endpoint remains the first zero of the physical
radius. We therefore monitor Eq.~(\ref{eq:globaladmiss}) globally. This
changes the interpretation of the first finite-time event.

\subsection{Body-force forcing and first interior contact}

We use the body-force forcing of Ref.~\onlinecite{Ramos1995},
\begin{equation}
 \frac{1}{\Fr(\tau)}=\frac{1}{\Fr_0}
 \left[1+A\sin(2\pi\St_g\tau)\right],
 \label{eq:bodyforcing}
\end{equation}
with zero initial phase and constant nozzle data. We use
$\Fr_0=10$, $\We=50$, $\theta_0=0$, $\beta=0$, $\Cpmax=1$,
$p_i^\ast(0)/p_e^\ast=1$, $A=0.5$ and $\St_g=0.5$. Because the sine starts
from its mean value, the initial field is the steady solution. There is no
first-order corner incompatibility. Differentiating the axial equation once
in time nevertheless shows a second-order mismatch at $(\tau,\eta)=(0,0)$:
the prescribed nozzle has $u_{\tau\tau}=0$ whereas the body-force term gives
$\partial_\tau(1/\Fr)|_{0}=2\pi\St_gA/\Fr_0=0.15708$. This explains the
small early residual layer near the nozzle and is one reason the off-grid
residual is reported on the explicit interior window of Sec.~\ref{sec:offgrid}.

The relevant event function is
\begin{equation}
   \Phi(\tau)=\min_{0\leq\eta<1}S(\tau,\eta).
   \label{eq:Phi}
\end{equation}
For each dense-output state we locate all real stationary points of the
Chebyshev interpolant of $S$ in $(-1,1)$ and minimise over those points and
the nozzle. The first sign change of $\Phi$ is then located in time by a
scalar root solve. This avoids introducing an arbitrary positive event
threshold. Table~\ref{tab:touchdown} shows convergence to
$t_c\simeq10.7538$ and $\eta_c\simeq0.9509$. At that instant the endpoint
is not close to losing transversality: $S(\tau_c,1)\simeq0.484$ on the
finest calculations. The scalar length $L\simeq13.141$ therefore still
tracks a simple endpoint zero, but the physical radius has acquired an
earlier zero at $z_c=L\eta_c\simeq12.496$.

\begin{table}[t]
\caption{First loss of global admissibility for body-force forcing
(\ref{eq:bodyforcing}) for $A=0.5$ and $\St_g=0.5$. The minimum is taken over
the complete Chebyshev interpolant on $0\leq\eta<1$, not only over
collocation nodes. The endpoint value $S(1)$ remains positive.}
\label{tab:touchdown}
\begin{ruledtabular}
\begin{tabular}{rrrr}
$N$ & $t_c$ & $\eta_c$ & $S(t_c,1)$\\
\colrule
24  & 10.7714840 & 0.939944 & 0.58551\\
32  & 10.7545050 & 0.947929 & 0.56845\\
48  & 10.7533145 & 0.951092 & 0.52215\\
64  & 10.7538308 & 0.950854 & 0.48631\\
96  & 10.7538402 & 0.950983 & 0.48421\\
128 & 10.7538189 & 0.950891 & 0.48472\\
160 & 10.7538262 & 0.950917 & 0.48433\\
\end{tabular}
\end{ruledtabular}
\end{table}

The small nonmonotone variation between the finest spectral values sets the
appropriate precision: we quote $t_c=10.7538$ and $\eta_c=0.9509$, not the
extra digits of any single discretisation. Temporal error is much smaller.
For $N=48$, DOP853 with relative tolerances $10^{-8}$, $10^{-10}$ and
$10^{-12}$ gives respectively $10.753314485734$, $10.753314485672$ and
$10.753314485672$, while Radau at relative tolerance $10^{-9}$ gives
$10.753314485714$.

\begin{figure}[t]
\includegraphics[width=\linewidth]{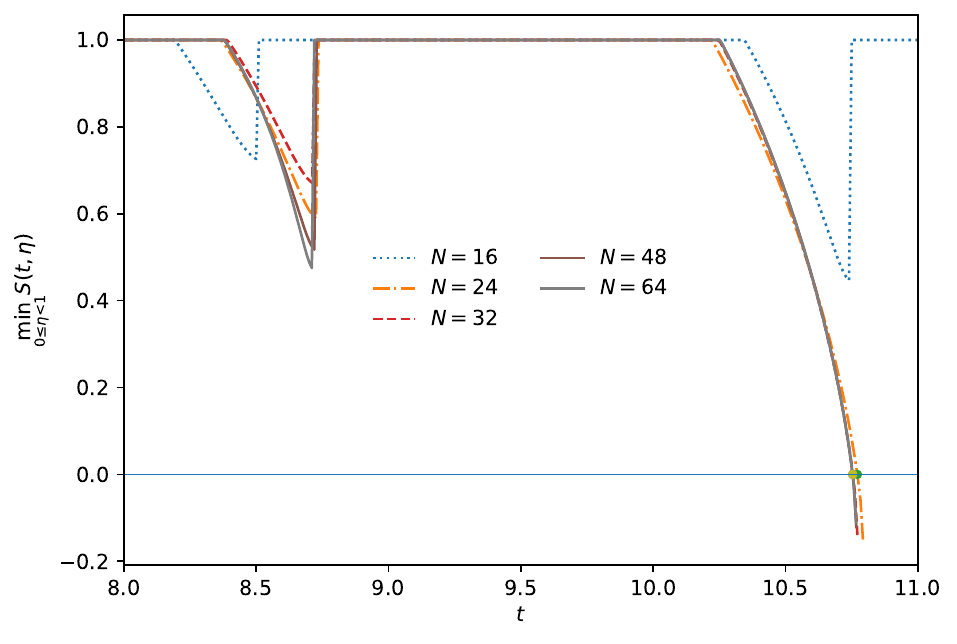}
\caption{Global admissibility measure $\Phi(t)$ of Eq.~(\ref{eq:Phi}) for
the body-force case of Table~\ref{tab:touchdown}. The coarse $N=16$
discretisation stays positive and continues to a periodic response, whereas
the refined solutions reach zero near $t_c=10.7538$. The earlier dip near
$t\simeq8.7$ remains positive.}
\label{fig:transversality}
\end{figure}

The independent finite-difference formulation of Sec.~\ref{sec:uw3} gives
the same limiting first-contact time without evolving $S$. There we monitor
$R/(1-\eta)$ at the uniform-grid interior points. Because the minimum is known only at the uniform-grid nodes, we define the
finite-difference event by $\min R/(1-\eta)=10^{-8}$. This small positive
threshold is used only to approximate the zero crossing in the independent
cross-check. The results in Table~\ref{tab:cross} converge from the
same side to the spectral zero-event value; the finest $M=1280$ result
differs from the $N=160$ spectral value by $1.6\times10^{-5}$. The $M=80$
mesh, like $N=16$, does not detect the event. Thus suppression of first
contact by coarse spatial resolution is not peculiar to the Chebyshev
representation.

\begin{table}[t]
\caption{Independent upwind finite-difference cross-check of the first
interior contact. The finite-difference event uses
$\min R/(1-\eta)=10^{-8}$ on the uniform-grid interior points. The
difference is relative to the $N=160$ spectral zero-event value
$t_c=10.7538262$.}
\label{tab:cross}
\begin{ruledtabular}
\begin{tabular}{rrr}
$M$ & $t_c$ & difference\\
\colrule
80   & not detected & \\
160  & 10.7595792 & $5.75\times10^{-3}$\\
320  & 10.7546248 & $7.99\times10^{-4}$\\
640  & 10.7539367 & $1.10\times10^{-4}$\\
1280 & 10.7538418 & $1.56\times10^{-5}$\\
\end{tabular}
\end{ruledtabular}
\end{table}

\subsection{Local structure and verification at contact}

At first contact the minimising point is an interior stationary point, so
$S(\tau_c,\eta_c)=S_\eta(\tau_c,\eta_c)=0$. The computed second derivative
is finite and positive. For $N=160$,
$S_{\eta\eta}=284.36$ and, because
$R=(1-\eta)S$, $R_{zz}=0.08082$. Hence
\begin{equation}
 R(\tau_c,z)=0.0404\,(z-z_c)^2+o((z-z_c)^2)
 \label{eq:quadraticcontact}
\end{equation}
locally. Moreover $S_\tau(\tau_c,\eta_c)=-5.26$, so the tangency is crossed
rather than merely touched in parameter time. This establishes the
quadratic character directly; it is not inferred from the vanishing of the
endpoint quantity $S(1)$.

The endpoint kinematics remain regular at the same instant. For $N=160$,
$\bar v(1)=-0.2384$, $\dot L=-4.556$ and
$L\bar v(1)/S(1)=-6.469$, all finite. Thus the formal
pole in Eq.~(\ref{eq:LdotS}) is irrelevant to the \emph{first} loss of
admissibility. Continuing the semidiscrete ODE beyond $t_c$ is numerically
possible for a short interval, but it no longer represents a single annular
sheet whose first closure is at $L$: $R$ becomes negative across an interior
interval before returning to the imposed endpoint zero. We therefore stop
the physical interpretation at $t_c$.

\begin{figure}[t]
\includegraphics[width=\linewidth]{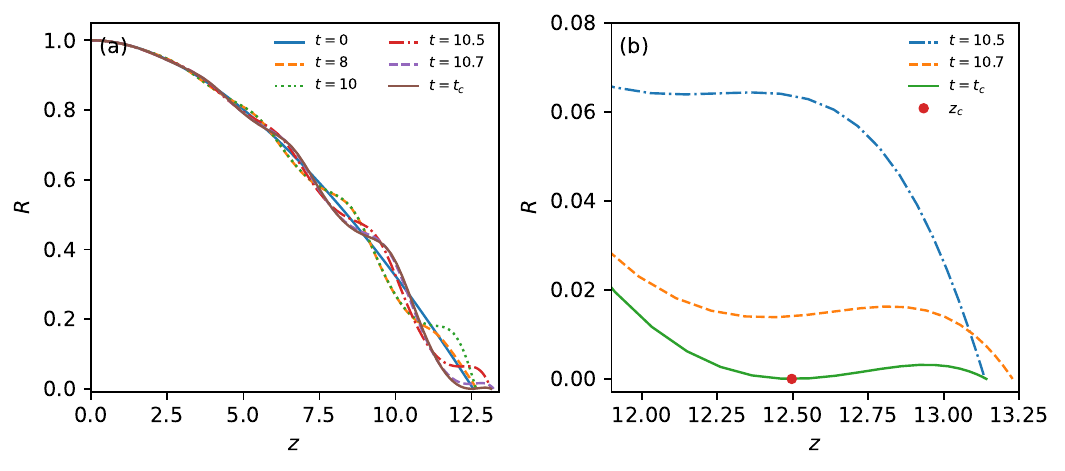}
\caption{Mean radius for the principal body-force calculation with $N=96$.
(a) Evolution over the full represented interval. (b) Detail near the
interior contact for $t=10.5$, $10.7$ and the first-contact time
$t_c=10.7538402$. At $t_c$ the radius is tangent to the axis at
$z_c=12.4968$ and becomes positive again before the separately imposed
endpoint zero at $L=13.1409$. Thus the endpoint remains transversal but is
no longer the first closure point.}
\label{fig:interface}
\end{figure}

The interior spatial residual supports the convergence assessment but is
not used as a stand-alone certificate. At $t=10.73383=t_c-0.02001$, the
maximum absolute off-grid residual on
$0.05\le\eta\le1-10^{-6}$ is $1.01\times10^{-2}$,
$1.54\times10^{-3}$, $2.80\times10^{-4}$ and $5.05\times10^{-6}$ for
$N=32$, 48, 64 and 96; the corresponding maximum equation-scaled residuals
are $9.60\times10^{-3}$, $1.30\times10^{-3}$,
$2.31\times10^{-4}$ and $6.24\times10^{-6}$. For $N=96$ the componentwise
absolute residuals $(m,S,u,\bar v)$ are
$(5.05\times10^{-6},5.91\times10^{-7},1.58\times10^{-7},9.74\times10^{-7})$,
and the largest Chebyshev-tail ratio is $1.21\times10^{-7}$. The algebraic
nozzle and endpoint residuals are on the order of round-off. Residual histories at
$t=0,2,5,8,10,10.5,10.7$ and $t_c-0.02$, together with all componentwise
values and scalings, are deposited with the revision data.

\subsection{Observation window and relation to earlier periodic responses}

For the same forcing frequency, $\St_g=0.5$, a smaller amplitude
$A=0.45$ produces no interior contact up to $t=100$ for either $N=48$ or
$N=64$. The integrations cover fifty forcing cycles. During these runs the
positivity safeguard used by the time integrator is never activated. The
last two phase-aligned cycles agree closely: the relative maximum difference
in $L$ is $3.50\times10^{-12}$ for $N=48$ and $8.37\times10^{-13}$ for
$N=64$, while the corresponding differences in $\Cpn$ are
$3.74\times10^{-11}$ and $1.66\times10^{-12}$. The endpoint and interior
positivity measures repeat to better than $6\times10^{-10}$. These results
support a periodic response over the finite observation window. They do not
prove stability for arbitrarily long times. Accordingly, for the tested
resolutions and $0\le t\le100$, we report only the numerical bracket
$0.45<A_{\rm onset}\le0.50$.

We also applied the global-contact test to the nozzle-forcing cases from the
submitted manuscript. In every case the first zero occurs in the interior
before the later endpoint event. For $N=64$, $a_f=0.25$ and $\St=0.1$ give
$t_c=11.1991$ with abrupt start-up and $t_c=31.5257$ with the two-cycle
ramp. For $\theta_0=15^\circ$, $a_f=0.1$ and $\St=0.1$, the corresponding
times are $21.5559$ and $31.8914$. These calculations show that the time of
first contact is sensitive to the start-up protocol. We therefore use them
as supporting checks rather than as the primary quantitative result.

The parameter set $A=0.5$, $\St_g=0.5$ was previously reported to produce a
periodic response.\cite{Ramos1995} In the present calculation, both the
spectral and finite-difference discretisations reach an interior contact
before such a long periodic continuation can be interpreted as a single
annular jet with its first closure at $L$. This comparison does not establish
why the earlier computation remained periodic. Numerical diffusion,
endpoint-closure drift and other implementation details are possible causes,
but distinguishing among them would require a direct reproduction of the
historical scheme and start-up procedure. Our conclusion is therefore
limited to the model solution computed here: it identifies the first loss of
global admissibility and the spatial resolution required to detect it.

\section{CONCLUSIONS}
\label{sec:conclusions}

The free end of the annular-jet model is governed by an algebraic closure
relation. Written as a semiexplicit system with endpoint speed $w=\dot L$ as
an algebraic variable, the spatially discretised problem has the regular
Hessenberg index-two structure while the terminal closure is transversal.
Differentiating that constraint is exact at the continuous level; numerical
drift arises only when a discrete method marches the differentiated relation
without also enforcing the original algebraic equation. For the membrane
limit $\beta=0$ the substitution $R=(1-\eta)S$ avoids this issue by
absorbing $R(1)=0$ identically. The same algebraic idea extends to finite
thickness through $R_i=R-\beta m/(2R)=(1-\eta)\Sigma$, for which
$2R^2-\beta m=2RR_i$; nevertheless, all unsteady calculations reported in
this paper are membrane calculations, and the finite-thickness numerical
results remain the separate steady comparison of Sec.~\ref{sec:thickness}.

The membrane equations, discretised by Chebyshev collocation, converge
geometrically for the smooth states considered here and reproduce published
steady data. The numerical solution is checked independently through
componentwise off-grid residuals, explicitly defined spectral tails,
temporal-tolerance convergence, and a finite-difference method based on
different variables and stencils. The enclosed-volume quadrature has also been corrected to an
$(N+2)$-point Gauss--Legendre rule, which is exactly sufficient for the
square of the represented radius; the numerical change is below the quoted
accuracy at the resolutions used.

The revised verification supports three conclusions, each with a
specific scope. First, at fixed $\Fr$, $\We$, $\theta_0$ and $\Cpn$, the steady
finite-thickness closure length decreases strictly with $\beta$ while the
closure level remains transversal, and the membrane limit has a
square-root correction. This is a parameter-space statement, not a universal
experimental law, and the disagreement with an earlier table only suggests,
rather than proves, a free-end implementation origin. Second, both the
nonlinear periodic calculation and the linear semidiscrete transfer function
show a pressure-coefficient resonance near $\St=0.093$, while the
convergence-length response is monotone over the verified globally
admissible band $0.005\le\St\le0.405$. Third, and most importantly, the
first finite-time failure in the principal body-force case is not loss of
transversality at the mapped endpoint. It is an interior tangential contact:
$t_c\simeq10.7538$, $\eta_c\simeq0.9509$, with
$S(t_c,1)\simeq0.484$. At contact the radius is locally quadratic in
$z-z_c$, and an independent upwind scheme converges to the same event.
Beyond it, the tracked endpoint is no longer the first physical closure and
the single-interval annular-jet representation has lost global admissibility.

The distinction between local endpoint regularity and global geometric
admissibility does not depend on the use of a spectral method. In any mapped
free-surface calculation whose endpoint is defined by a closure condition,
one should verify both that condition and the absence of an earlier interior
closure. Building the endpoint condition into the variables removes one
source of numerical drift, but the global geometric check remains necessary.

Two extensions are natural. One is the subcritical regime, where an upstream
capillary characteristic changes the admissible boundary data, as in planar
sheets.\cite{Benilov2019,Weinstein2019,Benilov2021,Chiatto2022} The other
is the dynamics after first interior contact. The present one-dimensional
single-interval model does not determine whether the physical sheet would
pinch off, reconnect or reopen; answering that question requires a model or
a resolved free-surface calculation that can change topology. The verified
pre-contact solution supplies a well-defined state from which such a study
can begin.

\begin{acknowledgments}
The author thanks Juan I.~Ramos for introducing him to annular liquid jets,
for a critical reading of the revised manuscript, and for helpful comments.
The author also thanks the two anonymous referees for their constructive
remarks. The author acknowledges the technical expertise and assistance provided
by the SCBI (Supercomputing and Bioinformatics Center) at the Universidad
de M\'alaga, including access to the Picasso Supercomputer. This work was
supported by the Universidad de M\'alaga, Spain, through project
PPRO-B4-2026-004 of its Plan Propio de Investigaci\'on, Transferencia y
Divulgaci\'on Cient\'ifica.
\end{acknowledgments}

\section*{AUTHOR DECLARATIONS}

\subsection*{Conflict of Interest}
The author has no conflicts to disclose.

\subsection*{Author Contributions}
\textbf{Francisco R. Villatoro:} Conceptualization; Formal analysis;
Investigation; Methodology; Software; Validation; Visualization;
Writing -- original draft; Writing -- review \& editing.

\section*{DATA AVAILABILITY}
The solvers, drivers and data for the revised manuscript are openly available
in Zenodo, version 1.1.0, at
\url{https://doi.org/10.5281/zenodo.23124655}, Ref.~\onlinecite{Villatoro2026data}.
The calculations in the original submission correspond to GitHub commit
\texttt{8b3672db1c\allowbreak dbf9a8f802\allowbreak ae116b631b\allowbreak 88a407c5fb}. The revised calculations,
including the global-admissibility, componentwise-residual, linear-response,
and finite-thickness closure-absorption audits, correspond to GitHub commit
\texttt{c46eaaa631\allowbreak fdb2515fd3\allowbreak 885e04a970\allowbreak 24bd3f3173}, archived as Zenodo v1.1.0.


\begin{thebibliography}{99}

\bibitem{Hoffman1980}
M.~A.~Hoffman, R.~K.~Takahashi, and R.~D.~Monson, ``Annular liquid jet
experiments,'' J. Fluids Eng. \textbf{102}, 344--349 (1980).

\bibitem{BairdDavidson1962b}
M.~H.~I.~Baird and J.~F.~Davidson, ``Annular jets---II. Gas
absorption,'' Chem. Eng. Sci. \textbf{17}, 473--480 (1962).

\bibitem{Sivakumar1997}
D.~Sivakumar and B.~N.~Raghunandan, ``A study on converging thin annular
jets,'' J. Fluids Eng. \textbf{119}, 923--928 (1997).

\bibitem{Sivakumar2002}
D.~Sivakumar and B.~N.~Raghunandan, ``Converging swirling liquid jets
from pressure swirl atomizers: Effect of inner air pressure,'' Phys.
Fluids \textbf{14}, 4389--4398 (2002).

\bibitem{Taylor1959}
G.~I.~Taylor, ``The dynamics of thin sheets of fluid. I. Water bells,''
Proc. R. Soc. London, Ser. A \textbf{253}, 289--295 (1959).

\bibitem{LancePerry1953}
G.~N.~Lance and R.~L.~Perry, ``Water bells,'' Proc. Phys. Soc. B
\textbf{66}, 1067--1072 (1953).

\bibitem{Clanet2001}
C.~Clanet, ``Dynamics and stability of water bells,'' J. Fluid Mech.
\textbf{430}, 111--147 (2001).

\bibitem{Clanet2007}
C.~Clanet, ``Waterbells and liquid sheets,'' Annu. Rev. Fluid Mech.
\textbf{39}, 469--496 (2007).

\bibitem{BairdDavidson1962}
M.~H.~I.~Baird and J.~F.~Davidson, ``Annular jets---I. Fluid dynamics,''
Chem. Eng. Sci. \textbf{17}, 467--472 (1962).

\bibitem{KihmChigier1990}
K.~D.~Kihm and N.~A.~Chigier, ``Experimental investigations of annular
liquid curtains,'' J. Fluids Eng. \textbf{112}, 61--66 (1990).

\bibitem{Kendall1986}
J.~M.~Kendall, ``Experiments on annular liquid jet instability and on the
formation of liquid shells,'' Phys. Fluids \textbf{29}, 2086--2094
(1986).

\bibitem{LeeWang1986}
C.~P.~Lee and T.~G.~Wang, ``A theoretical model for the annular jet
instability,'' Phys. Fluids \textbf{29}, 2076--2085 (1986).

\bibitem{SiamasJiang2009}
G.~A.~Siamas, X.~Jiang, and L.~C.~Wrobel, ``Direct numerical simulation
of the near-field dynamics of annular gas--liquid two-phase jets,'' Phys.
Fluids \textbf{21}, 042103 (2009).

\bibitem{Ramos1992ZAMM}
J.~I.~Ramos, ``Annular liquid jets: Formulation and steady state
analysis,'' ZAMM Z. Angew. Math. Mech. \textbf{72}, 565--589 (1992).

\bibitem{RamosFalgueras1991}
J.~I.~Ramos and J.~Falgueras, ``Response of annular liquid jets to mass
loading,'' Comput. Mech. \textbf{9}, 1--16 (1991).

\bibitem{Ramos1992AMM}
J.~I.~Ramos, ``Adaptive block-implicit methods for annular liquid jets,''
Appl. Math. Modelling \textbf{16}, 464--475 (1992).

\bibitem{RamosFalgueras1992}
J.~I.~Ramos and J.~Falgueras, ``Oscillating annular liquid membranes,''
Arch. Appl. Mech. \textbf{62}, 43--52 (1992).

\bibitem{Ramos1995}
J.~I.~Ramos, ``The effects of fluctuating body forces on annular liquid
jets,'' Arch. Appl. Mech. \textbf{65}, 548--563 (1995).

\bibitem{RamosIJNMF1995}
J.~I.~Ramos, ``Fluid dynamics of slender, thin, annular liquid jets,''
Int. J. Numer. Methods Fluids \textbf{21}, 735--761 (1995).

\bibitem{RamosIJNMHFF1992}
J.~I.~Ramos, ``Mass absorption by annular liquid jets: III. Numerical
studies of jet collapse,'' Int. J. Numer. Methods Heat Fluid Flow
\textbf{2}, 21--36 (1992).

\bibitem{Ramos1993}
J.~I.~Ramos, ``Domain-adaptive finite difference methods for collapsing
annular liquid jets,'' Comput. Mech. \textbf{11}, 28--64 (1993).

\bibitem{Ramos1996}
J.~I.~Ramos, ``Upward and downward annular liquid jets: Conservation
properties, singularities, and numerical errors,'' Appl. Math. Modelling
\textbf{20}, 440--458 (1996).

\bibitem{Brunet2004}
P.~Brunet, C.~Clanet, and L.~Limat, ``Transonic liquid bells,'' Phys.
Fluids \textbf{16}, 2668--2678 (2004).

\bibitem{LhuissierVillermaux2012}
H.~Lhuissier and E.~Villermaux, ``Crumpled water bells,'' J. Fluid Mech.
\textbf{693}, 508--540 (2012).

\bibitem{Benilov2019}
E.~S.~Benilov, ``Oblique liquid curtains with a large Froude number,'' J.
Fluid Mech. \textbf{861}, 328--348 (2019).

\bibitem{Weinstein2019}
S.~J.~Weinstein, D.~S.~Ross, K.~J.~Ruschak, and N.~S.~Barlow, ``On
oblique liquid curtains,'' J. Fluid Mech. \textbf{876}, R3 (2019).

\bibitem{Benilov2021}
E.~S.~Benilov, ``Paradoxical predictions of liquid curtains with surface
tension,'' J. Fluid Mech. \textbf{917}, A21 (2021).

\bibitem{DellaPia2023}
A.~Della~Pia, M.~G.~Antoniades, E.~S.~Ioannidis, Z.~A.~Wejko,
N.~S.~Barlow, M.~Chiatto, S.~J.~Weinstein, and L.~de~Luca, ``On the
shapes of liquid curtains flowing from a non-vertical slot,'' J. Fluid
Mech. \textbf{974}, A18 (2023).

\bibitem{Chiatto2022}
M.~Chiatto and A.~Della~Pia, ``Natural frequency discontinuity of
vertical liquid sheet flows at transcritical threshold,'' J. Fluid Mech.
\textbf{945}, A32 (2022).

\bibitem{DellaPia2021}
A.~Della~Pia, M.~Chiatto, and L.~de~Luca, ``Receptivity to forcing
disturbances in subcritical liquid sheet flows,'' Phys. Fluids
\textbf{33}, 032113 (2021).

\bibitem{DellaPia2020}
A.~Della~Pia, M.~Chiatto, and L.~de~Luca, ``Global eigenmodes of thin
liquid sheets by means of volume-of-fluid simulations,'' Phys. Fluids
\textbf{32}, 082112 (2020).

\bibitem{Finnicum1993}
D.~S.~Finnicum, S.~J.~Weinstein, and K.~J.~Ruschak, ``The effect of
applied pressure on the shape of a two-dimensional liquid curtain falling
under the influence of gravity,'' J. Fluid Mech. \textbf{255}, 647--665
(1993).

\bibitem{Torsey2021}
B.~Torsey, S.~J.~Weinstein, D.~S.~Ross, and N.~S.~Barlow, ``The effect of
pressure fluctuations on the shapes of thinning liquid curtains,'' J.
Fluid Mech. \textbf{910}, A38 (2021).

\bibitem{Girfoglio2017}
M.~Girfoglio, F.~De~Rosa, G.~Coppola, and L.~de~Luca, ``Unsteady critical
liquid sheet flows,'' J. Fluid Mech. \textbf{821}, 219--247 (2017).

\bibitem{Ramos1997}
J.~I.~Ramos, ``Analysis of annular liquid membranes and their
singularities,'' Meccanica \textbf{32}, 279--293 (1997).

\bibitem{Villatoro2026data}
F.~R.~Villatoro, ``Annular liquid jets: solvers and data for exact
constraint absorption and finite-time loss of global admissibility,'' Zenodo,
version 1.1.0 (2026), \url{https://doi.org/10.5281/zenodo.23124655}.

\end{thebibliography}
\end{document}